\documentclass[a4paper]{article}

\usepackage{INTERSPEECH2020}
\usepackage[utf8]{inputenc}
\usepackage[tone]{tipa}
\usepackage{bbold}
\DeclareMathOperator{\logsumexp}{logsumexp}
\usepackage{stackrel}

\title{Autosegmental Neural Nets: Should Phones and Tones be Synchronous or Asynchronous?}
\name{Jialu Li, Mark Hasegawa-Johnson}
\address{Department of Electrical and Computer Engineering $\&$ Beckman Institute, University of Illinois}
\email{\{jialuli3, jhasegaw\}@illinois.edu}

\begin{document}

\maketitle
\begin{abstract}
  Phones, the segmental units of the International Phonetic Alphabet (IPA), are used for lexical distinctions in most human languages; Tones, the suprasegmental units of the IPA, are used in perhaps 70\%.  Many previous studies have explored cross-lingual adaptation of automatic speech recognition (ASR)  phone models, but few have explored the multilingual and cross-lingual transfer of synchronization between phones and tones. In this paper, we test four Connectionist Temporal Classification (CTC)-based acoustic models, differing in the degree of synchrony they impose between phones and tones.  Models are trained and tested multilingually in three languages, then adapted and tested cross-lingually in a fourth. Both synchronous and asynchronous models are effective in both multilingual and cross-lingual settings. Synchronous models achieve lower error rate in the joint phone+tone tier, but asynchronous training results in lower tone error rate.
\end{abstract}

\section{Introduction}

In most of the world's languages (possibly as many as 70\%~\cite{Yip2002}), the meaning of a word depends on both phones and tones.  Phones are called segmental because their acoustic cues occur in and around discrete temporal segments.  Tones are called suprasegmental because each tone may be aligned over one or more segments.  In Mandarin, for example, tones are canonically synchronized with the vowel and coda consonant of each syllable~\cite{Xu1998}, but may influence the onset of the following syllable~\cite{Xu1999}, and may be adopted, in an apparently rule-driven manner, as the pitch of a following neutral-tone syllable~\cite{Chen2000}.  Similar rightward spreading occurs in many, but not all, tone languages~\cite{Hyman1974}.  In many languages, rightward spreading of a tone is not blocked by intervening vowels, consonants, or even syllables, but only by the intervention of another tone, suggesting that tones and phones are ``autosegmental'' (communicated as loosely-related segmentations of the time axis)~\cite{Goldsmith1975,Goldsmith1976}.

Most hidden Markov model-based (HMM-based) ASR in non-tonal languages uses one HMM per phone or triphone~\cite{Lee90a}.  HMM-based ASR for tonal languages, by contrast, may use one HMM per final~\cite{Lin93}, per complete syllable~\cite{Lee2002}, or per sequence of two to three syllables~\cite{Qian2007} so that the canonical domain of the lexical tone can be learned by the HMM.  Localizing lexical tone on the vowel of each syllable is possible in a deep neural network (DNN)-HMM hybrid, apparently because the DNN captures sufficient acoustic context~\cite{Can_viet_tone,viet_tone_work}.

End-to-end neural ASR, trained using CTC~\cite{CTC}, can sidestep the tone-to-phone alignment problem by generating characters, rather than phones, as the output~\cite{Hannun2014}.
In a CTC system with character outputs, however, it is difficult to share data for multilingual~\cite{Vesely2012} or cross-lingual~\cite{Byrne2000} ASR.
Proposed solutions have included separate softmax tiers for the character set of each language~\cite{transfer_work1,transfer_work2,transfer_work4}, or the generation of phone strings instead of characters as the output of the CTC~\cite{ctc_work_phone,ctc_work_phone2,ctc_work_phone3}, or the use of both methods, in a multi-task learning framework, with one output tier generating phones, while another generates characters~\cite{transfer_work3}.

Mixed tones and phones using CTC have been demonstrated for Mandarin~\cite{Qu2017} and for two under-resourced tonal languages~\cite{Adams2017}, but there have been few studies (if any) about multilingual or cross-lingual modeling of tone-marked-phones using CTC.   Different tone languages seem to lend themselves to different temporal domains for the tone, e.g., Mandarin benefits from tone-marked finals~\cite{Zhang2019} or syllables~\cite{Qu2017,Zhao2018}, while ASR in other tone languages has used tone-marked vowels~\cite{Adams2017}.  There is also some disagreement about whether tones and phones should be modeled jointly, or separately.  For example, monolingual systems trained for the under-resourced tonal languages Na and Chatino found that both phone error rate (PER) and tone error rate (TER) are lower, in a CTC-based recognizer, if the phones and tones are modeled together, rather than separately, unless the recognizer is trained using at least 120 minutes of training data per language~\cite{Adams2017}.  With at least 120 minutes of data, the results were mixed: the joint system gave lower TER but higher PER in Na, but the opposite result in Chatino.

This paper performs Multilingual and Cross-lingual recognition of phones and tones using end-to-end neural networks trained using CTC.  Trained recognizers are tested on languages within the training set (Multilingual test), and adapted to a language with minimal adaptation data (Cross-lingual test).   Four different systems are tested.  The first system generates phone-marked tones as its output, similar to the joint transcription model of~\cite{Adams2017}.  The second system has two separate output tiers, similar to the phones and characters of~\cite{transfer_work3}, but containing, instead, phones and tones.  The third system combines the first two, with three output tiers.  The fourth system is similar to the third, but standardizes tones across languages by forcing every tone, in every language, to have exactly two pitch targets.

The paper is organized as follows:
Section \ref{model} introduces the CTC-based acoustic models and cross-language adaptation methods in detail. Section \ref{experiment} describes datasets and experimental methods. Section \ref{discussions} provides results for each system, followed by analysis of multilingual and cross-lingual phone and tone modeling effects. Finally, Section \ref{conclusions} concludes.

\section{Model}
\label{model}





\begin{figure}
  \centering
   \includegraphics[width=1.01\linewidth]{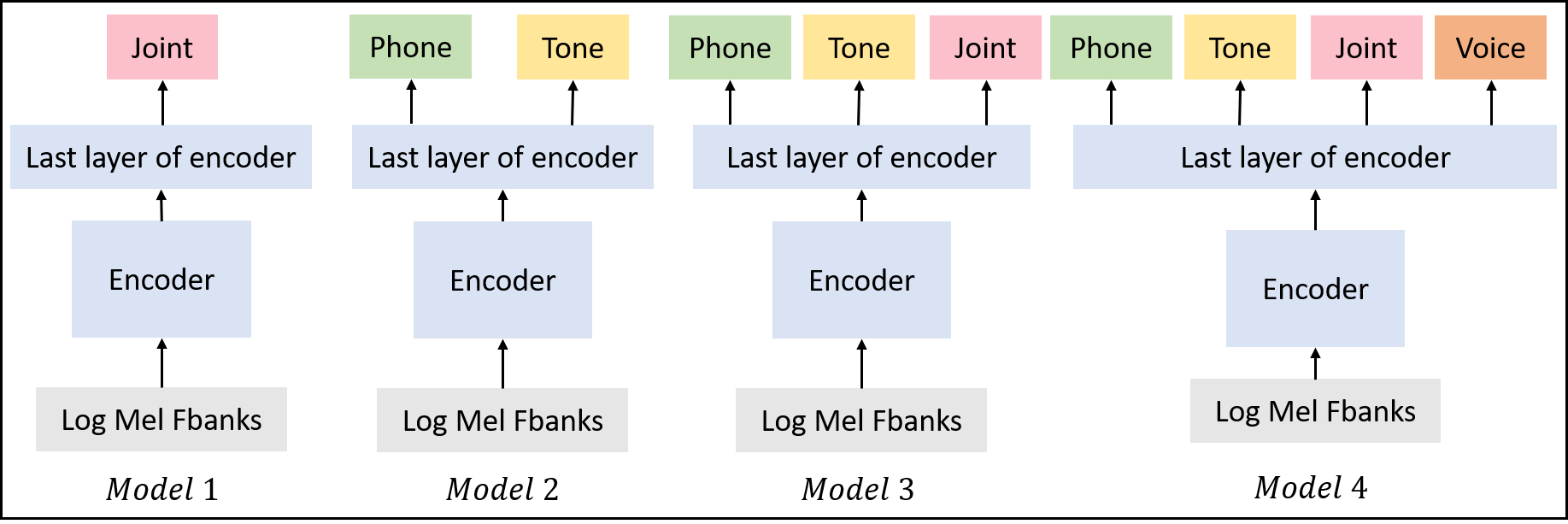}
\caption{CTC-based acoustic model with different multi-task learning tiers. }
  \label{fig:model}
\end{figure}


Four different end-to-end multilingual ASR systems were trained, using a CTC training criterion (Figure~\ref{fig:model}).  All four systems used language-independent encoder networks (bLSTM$\times$3+one fully-connected layer), followed by a language-dependent softmax layer.  All four  systems were designed to learn a mapping from acoustic inputs ($x=[x_1,\ldots,x_N]$) to sequences of phonetic label outputs arranged in one or more tiers.  Let the reference label sequence in the $t^{\textrm{th}}$ tier be $\pi^{(t)}=[\pi_1^{(t)},\ldots,\pi_m^{(t)}]$, where $\pi_j^{(t)}\in\mathcal{A}^{(t,l)}$ is a symbol in the $t^{\textrm{th}}$ sub-alphabet of the $l^{\textrm{th}}$ training language, and the length $m$ is different in different tiers.  Our systems are trained using a standard CTC training criterion~\cite{CTC} in each tier, which can be written as
\begin{equation}
  \mathcal{L}^{(t)}=-\stackrel[y^{(t)}:\mathcal{B}(y^{(t)})=\pi^{(t)}]{}{\logsumexp}\sum_{i=1}^n \log p(y_i^{(t)}|x_i),
  \label{eq:ctc_loss}
\end{equation}
where $y^{(t)}=[y_1^{(t)},\ldots,y_n^{(t)}]$, $y_i^{(t)}\in\mathcal{A}^{(t,l)}\cup\left\{b\right\}$, $b$ is the blank symbol, and the operation $\mathcal{B}(y^{(t)})$ eliminates sequential duplicates, then eliminates blanks~\cite{CTC}.  The four systems shown in Figure~\ref{fig:model} differ only in the number and alphabets of their output tiers.

\begin{table*}
  \caption{Lexical tones and glottal phones that are part of the phoneme inventories of the four tonal languages used in this study: Mandarin, Cantonese, Vietnamese, and Lao.}
  \label{tab:tone}
  \centering
  \begin{tabular}{llllll|ll}
    \toprule
    Man & Descriptions & Can & Descriptions& Viet & Descriptions& Lao & Descriptions       \\
    \midrule
    a\tone{55} & high & a\tone{55} & high & a\tone{33} & mid & a\tone{11} & low \\
    a\tone{33}\tone{55} & mid-rising & a\tone{33} & mid & a\tone{33}\tone{55} & mid-rising & a\tone{33} & mid \\ 
    a\tone{22}\tone{11}\tone{44}    & low-dipping & a\tone{22} & low & a\tone{33}\tone{22}\textglotstop &mid-falling, glottalized & a\tone{55} & high\\
    a\tone{55}\tone{11}    & falling & a\tone{33}\tone{55} & mid-rising & a\tone{22}\tone{11}\textipa{h} & mid-falling, breathy & a\tone{11}\tone{33} & low-rising\\
    & &a\tone{22}\tone{11} & low-falling & a\tone{22}\tone{11}\tone{22} & low-falling & a\tone{55}\tone{33} & high-falling\\
    & &a\tone{11}\tone{33} & low-rising & a\tone{33}\textglotstop\tone{55} & mid-rising, stopped & a\tone{33}\tone{11} & mid-falling\\  
    & &&&  & & \textglotstop & glottal stop\\  
    & &&&  & & \textipa{h} & glottal fricative\\  
    \bottomrule
  \end{tabular}
\end{table*}

Model 1 has one output tier per training language, whose alphabet includes all consonant phonemes and all tone-marked vowel phonemes of the language.  For example, the Mandarin-language softmax layer contains five variants of the vowel \textipa{[a]}: the vowel with neutral tone, and the vowel with four different lexical tones, as shown in the first column of Table~\ref{tab:tone}.

Model 2 has two output tiers: phones and tones. The alphabet of the phone tier in each language is the set of its segmental phonemes.  The alphabet of the tone tier is a universal phonetic tone inventory described in Section\ref{tone_rule}.  Model 3 has three output tiers, exactly equal to the joint tier of model 1, and the phone and tone tiers of model 2.  Model 4 standardizes the tone transcripts across languages, and extracts a separate voice quality tier, as described in Section~\ref{tone_rule}.  All four systems are trained using a multi-task loss function with equal weights for each tier, i.e.,
$\mathcal{L}=\sum_{t} \mathcal{L}^{(t)}$.

\subsection{Cross-lingual phone transfer}
\label{phone_rule}

Two types of experiments were performed in this study: Multilingual ASR (training and testing on different speech data from the same set of training languages), and Cross-lingual ASR (the fully trained model is adapted using limited data, then tested on the adaptation language).  In order to initialize the fully-connected layers for the adaptation language, we adopted a strategy similar to~\cite{Byrne2000,Scharenborg2017,transfer_work1,Li2020a}, based on knowledge-based cross-lingual mapping of IPA \cite{IPA1999} symbols.  The softmax layer of the adaptation language is initialized as follows: denote the dense layer weight matrix as $W^{(t,l)} \in \mathbf{R}^{(1+|\mathcal{A}^{(t,l)}|) \times d}$, where $\mathcal{A}^{(t,l)}$ is the alphabet of tier $t$ in language $l$, $1$ is the blank character, and $d$ is the dimension of the hidden layer. For a target phone $k$ in the adaptation language, if $k$ exists in any training language, then the average over the corresponding entries of all training language weight matrices is used to initialize the adaptation language.  If phone $k$ exists in no training language, then it is initialized, if possible, using a phone $k'$ that is equal to $k$ plus a diacritic, e.g., the phone \textipa{[a]} could be initialized by \textipa{[a:]}.  If there is no such extension, then finally, $k$ is initialized by a phone $k'$ that is most similar according to the consonant or vowel features of the IPA chart~\cite{IPA1999}, e.g., the vowel \textipa{[7]} could be initialized by \textipa{[o]}. Similarly, for the joint tier, the closest phone is first located, then among the candidate tone-marked versions of that phone, the one with the closest tone is located, e.g., \textipa{[u:\tone{11}\tone{33}]} could be initialized using \textipa{[u:\tone{33}\tone{55}]}, and \textipa{[\textscripta o\tone{11}\tone{33}]} could be initialized using \textipa{[\textscripta \textupsilon\tone{11}\tone{55}]}.   Once the closest phones among training languages have been identified, then the corresponding weights of the adaptation language are initialized as
\begin{equation} 
    \label{eq:weight_adapt}
    W_{k}^{(t,l)}=\left\{
    \begin{array}{ll}
         \frac{\sum_{l'=1}^n W_{k}^{(t,l')}}{\sum_{l'=1}^n\mathbb{1}_{k\in \mathcal{A}^{(t,l')}}}
          &\text{ if } \sum_{l'=1}^n\mathbb{1}_{k\in \mathcal{A}^{(t,l')}}\neq 0\\
         \frac{\sum_{l'=1}^nW_{k'}^{(t,l')}}{\sum_{l'=1}^n\mathbb{1}_{k'\in \mathcal{A}^{(t,l')}}} & \text{else}
    \end{array}
    \right.
\end{equation}
where $n$ is the number of training languages, $W_{k}^{(t,l)}$ is the weight vector for phone $k$ in tier $t$ of language $l$, $\mathcal{A}^{(t,l)}$ is the corresponding alphabet, and $\mathbb{1}$ denotes the identity function.

\subsection{Cross-lingual tone transfer}
\label{tone_rule}

Lexical tone is suprasegmental: it is not necessarily time-aligned with any single phone segment.  Standard IPA transcription methods list a lexical tone as a sequence of tone targets following the vowel, but it is not clear that synchronizing the tones in this way helps ASR.  In order to explore possible asynchrony  between tones and phones, Models 2, 3, and 4 use separate tone tier outputs.  In order to further reduce synchronization requirements, the alphabet of the tone  tier is not linked to the particular lexical tone inventory of each language: instead, the alphabet of this tier is language-independent, and consists of the five distinct IPA tone targets (extra high(\tone{55}), high(\tone{44}), mid(\tone{33}), low(\tone{22}), and extra low(\tone{11})), the symbol $\langle\text{neutral}\rangle$ as a placeholder for a syllable with neutral or unmarked tone, and a $\langle\text{boundary}\rangle$ symbol to mark syllable boundaries.  Models 2 and 3, but not Model 4, augment this alphabet with the symbols [\textglotstop] and \textipa{[h]}, in order to correctly label the glottalized and breathy tones of Vietnamese.

Model 4 attempts some degree of cross-language standardization, in both the length and content of the tone targets in each syllable.  Tone-tier training transcripts for Model 4 were normalized prior to training and testing, so that each syllable corresponds to exactly three characters: two tones, and a syllable boundary.  Lexical tones that are canonically transcribed with three IPA symbols, like Mandarin tone 3 (\tone{22}\tone{11}\tone{44}), were truncated (\tone{22}\tone{11}).  Tones that are usually transcribed with one target, including neutral tones and, e.g., Mandarin tone 1 (\tone{55}), were reduplicated (\tone{55}\tone{55}).  Voice quality symbols in the canonical tone descriptions of Vietnamese ([\textglotstop] and~\textipa{[h]}) were moved to a new voice quality tier, as were the corresponding phone segments in Lao.  In order to maintain structure in the voice-quality transcripts, each syllable received at least one voice quality marker: either [\textglotstop], or \textipa{[h]}, or a new modal-voicing symbol, $\langle\text{modal}\rangle$.  The resulting tier alphabets for Model 4 are $\mathcal{A}^{(tone)}=\{\text{\tone{11}},\text{\tone{22}},\text{\tone{33}}, \text{\tone{55}}, \langle\text{neutral}\rangle,\langle\text{boundary}\rangle\}$ and $\mathcal{A}^{(voice)}=\{\text{\textglotstop},\text{\textipa{h}},\langle\text{boundary}\rangle, \langle\text{modal}\rangle\}$.

\section{Experimental methods}
\label{experiment}


Sources of data, and quantities used for training, development, and test sets are listed in Table \ref{tab:data}. In order to test Cross-lingual ASR, the Lao dataset was artificially restricted to just 1 hour for adaptation, 1 hour for development, and 1 hour for testing.

\begin{table}
  \caption{Sources of data, and quantities (hours) used for Multilingual and Cross-lingual training, development, and testing.}
  \label{tab:data}
  \centering
  \setlength{\tabcolsep}{4.0pt}
  \begin{tabular}{l||c|c|ccc}
    \toprule
    Setting & Language & Source &Train & Dev& Test\\
    \midrule
    Multi &Mandarin & HUB4-NE&26.50 & 1.46 & 1.43\\
    &Cantonese &BABEL & 31.08 & 1.56 & 1.84\\
    &Vietnamese &BABEL& 18.24 & 1.54 & 1.39\\
    \midrule\midrule
    Cross&    Lao & BABEL& 1&1&1\\
    \bottomrule
  \end{tabular}
  
\end{table}
BABEL speech corpora consist of conversational and scripted data for each language; we used scripted data only because of its better audio quality. We found that conversational speech data often contains noise and long silences. 

All experiments were performed using extracted 40-dimensional log Mel filterbank features,
computed using the python speech features library~\cite{python_speech_features}, with a 25ms Hamming window and 10ms shift.  Each feature dimension was $Z$-normalized per speaker. 
One additional experiment was performed with Model 1, in which its input feature vector was augmented by a fundamental frequency measurement (F0), because F0 has been shown to reduce ASR error rates for tonal languages \cite{f0_work1,f0_work2}.  F0 was extracted from the same 25ms windowed frame, converted from Hertz to Mel scale, then appended to the 40-dimensional log Mel features. Model 1 was chosen for augmentation because it gave the lowest joint error rate in the Cross-lingual train/test condition, as described in Section~\ref{discussions}.

IPA phone transcripts were created for each language using the LanguageNet Grapheme-to-Phoneme (G2P) transducers \cite{languageNet2} implemented in Phonetisaurus~\cite{G2P} to generate IPA-based phonetic transcripts for each utterance. Vowels usually have tones associated with them, and consonants often don't have tones associated with them. We extracted each phone and its corresponding tone letters respectively as described in subsection \ref{phone_rule} and subsection \ref{tone_rule} to prepare for multitask learning in the acoustic modeling. 

Models were implemented using the eXtensible Neural Machine Translation toolkit \cite{xnmt}. Three layers of pyramidal Bi-directional Long-Short Term Memory (pBLSTM) are used as the encoder. The hidden dimension of the fully-connected layer is $d=512$; the input and hidden dimensions of the LSTM layer are 1024 and 256. The optimizer is Adadelta, with a learning rate of 0.004, and with early stopping using the development set to choose the best model. Decoding used a beam search with language modeling to obtain the best results on test set; beam width is 25 and the language modeling coefficient is 0.1.

For Cross-lingual adaptation, softmax output layers for Lao were (1)  initialized as described in Section~\ref{phone_rule}, (2) retrained without updating the adapted encoder, then (3) fine-tuned together with the encoder until convergence.

A Monolingual system was trained and tested as a baseline for Cross-lingual adaptation in Lao.  The Monolingual baseline used the same architectures as the Cross-lingual systems, but when trained on 1 hour of data using the same number of parameters as the Cross-lingual system, it failed to converge.  In order to achieve error rates below 100\%, therefore, the parameter count of the Monolingual system was greatly reduced. The input and hidden dimensions of the LSTM layers, and the hidden layer dimension of the fully-connected layer, are reduced as necessary to minimize development-set error rates, resulting in dimensions of 2--4 nodes each.

\section{Results and discussion}
\label{discussions}

\begin{table}
  \caption{Phone error rates (PER), tone error rates (TER), joint phone and tone error rates (JER), and voice quality error rates (VER) in the Multilingual (trained and tested on different speech data from the same three languages), Cross-lingual (adapted using one hour), and Monolingual (trained using one hour) settings, in percent.  M1+F0=Model 1 with both Mel filterbank and F0 input features.  Lowest number in each column is bold.}
  \label{tab:multitier}
  \centering
  \setlength{\tabcolsep}{5.0pt}
  \begin{tabular}{lc||c|c|c||c||c}
    \toprule
    &&\multicolumn{3}{|c||}{Multilingual}&Cross&Mono\\
    & & Man & Can & Viet & Lao & Lao \\
    \midrule
    JER&Model1 &55.73 &45.95 &53.45 & 54.36 &83.81\\
    &Model3 &61.07 &45.91 &53.37 & 69.32 &{\bf 82.36}\\
    &Model4 &60.22 &46.13 &53.49 & 81.72 &84.93\\    
    &M1+F0 &{\bf 55.35} & {\bf 40.31} & {\bf 48.91} & {\bf 53.26} & - \\
    \midrule
    PER&Model2 &59.88 &47.02 &55.51 & {\bf 57.69} &{\bf 90.05}\\
    &Model3 &52.59 &{\bf 39.97} &49.69 & 60.88 &90.53\\
    &Model4 &{\bf 51.60} &40.34 &{\bf 49.04} & 77.97 &90.74\\
    \midrule
    TER&Model2 &58.32 &43.80 &48.05 & {\bf 44.34} &{\bf 79.01}\\
    &Model3 &62.34 &39.19 &44.59 & 46.88 &82.52\\
    &Model4 &{\bf 52.09} &{\bf 39.02} &{\bf 33.91} & 68.04 &92.53\\
    \midrule
    VER&Model4 &- &- &37.08 & 75.11 &90.42\\    
    \bottomrule
  \end{tabular}
\end{table}

Table~\ref{tab:multitier} shows error rates of all four models, and of Model 1 with both Mel filterbank and F0 inputs (M1+F0).  Three experimental settings are distinguished: Multilingual (trained using 75.82 hours of Mandarin, Cantonese, and Vietnamese, tested on a different 4.66 hours in the same languages), Cross-lingual (adapted using 1 hour of Lao, tested using 1 hour of Lao), and Monolingual (trained using 1 hour, tested using 1 hour).  Cross-lingual training is better than Monolingual training, for all Models, and for all error metrics, but closer analysis  reveals striking differences between the different test conditions. {\bf MULTILINGUAL:} Joint phone+tone error rate (JER) is either lowest for Model 1 (Mandarin and Lao) or roughly comparable across Models 1, 3, and 4 (Cantonese and Vietnamese), but in all four languages, JER is significantly reduced by adding F0 to the input feature  vector (M1+F0).  Phone error rate (PER)  and tone error rate (TER) are much worse in Model 2 than in Models 3 and 4. For Mandarin, JER, PER and TER are relatively higher for all four models. This is perhaps due to the noisy speech collected while reporters were interviewing outdoor and some code-switching utterances that the models failed to generate correct phonemes. {\bf MONOLINGUAL:} TER is lowest in Model 2, suggesting that lexical tones in Lao may be best learned in  isolation (without the joint tier), and JER is lowest in Model 3, suggesting that the joint phones+tones tier may be best learned in combination with the tones-only tier.  {\bf CROSS-LINGUAL:} the smaller the model is, the better, within the limits of the optimized-parameter-count  systems shown in Table~\ref{tab:multitier}.  JER is lowest in Model 1, while PER and TER are lowest in Model 2.

Model 4 has the lowest TER in the Multilingual setting, but its superiority may be caused by its lower cardinality: as described in Section~\ref{tone_rule}, the tone tier of Model 4 has an output alphabet with only 6 output symbols (plus blank), while those of Models 2 and 3 both have output alphabets containing 9 output symbols (plus blank).  Even if the superiority of Model 4 is discounted, however, the key finding of the TER section in Table~\ref{tab:multitier} is unchanged.  Model 3 has lower TER than Model 2 in the Multilingual case, but not in Monolingual or Cross-lingual.  The key finding remains, therefore, that multi-task training of the tone tier together with a joint tier improves TER in the Multilingual setting, but not in the Monolingual or Cross-lingual settings.


\begin{table}
  \caption{Consonant error rates (CoER), vowel error rates (VoER), phone error rates (PER), and tone error rates (TER) computed from joint tier in Models 1, 3 and 4 and phone tier in Model 2. Lowest number in each column is bold only if lower than the corresponding best result in Table~\ref{tab:multitier}.}
  \label{tab:multitier_breakdown}
  \centering
  \setlength{\tabcolsep}{5.0pt}
  \begin{tabular}{lc||c|c|c||c}
    \toprule
    &&\multicolumn{3}{|c||}{Multilingual}&Cross\\
     && Man & Can & Viet & Lao \\
    \midrule
    CoER&Model1 &46.34 &50.21 &48.23 & {\bf 39.43}\\
    &Model2  &64.22 &68.41 &52.94 & 72.95 \\
    &Model3 &52.53 &53.78 &51.47 & 48.77 \\
    &Model4 &52.00 &55.37 &48.84 & 60.77 \\
    &M1+F0 &{\bf 46.12} &{\bf 46.61} &{\bf 45.19} & 41.77\\
    \midrule
    VoER & Model1 &49.11 &33.94 &54.73 & 61.67\\
    &Model2 &53.00 &39.75 &64.04 & {\bf 51.78} \\
    &Model3 &53.72 &31.52 &54.85 & 76.81 \\
    &Model4 &56.84 &32.25 &65.43 & 91.46 \\
    &M1+F0 &{\bf 48.93} &{\bf 28.33} &{\bf 50.00} & 57.67\\
    \midrule
    PER & Model1 &{\bf 48.61} &41.12 &56.33 & 50.98\\
    &Model3 &53.80 &40.11 &57.97 & 66.55 \\
    &Model4 &55.65 &40.30 &59.39 & 77.78 \\    
    &M1+F0 &54.81 &{\bf 34.72} &51.42 & {\bf 48.62}\\
    \midrule
    TER & Model1 &55.35 &43.02 &51.42 & 68.31\\
    &Model3 &58.14 &39.38 &51.65 & 77.99 \\
    &Model4 &54.79 &40.16 &53.08 & 91.83 \\
    &M1+F0 &54.41 &{\bf 37.80} &49.18 &  67.12\\


    \bottomrule
  \end{tabular}
  
\end{table}

Model 1's superior JER, in the Cross-lingual case, suggests an experiment in which PER and TER are measured using the phone and tone symbols produced by Model 1's joint output tier.  Table~\ref{tab:multitier_breakdown}  shows the PER and TER of phones and tones extracted from the joint output tiers of Models 1, 3, and 4.  Table~\ref{tab:multitier_breakdown} also shows the consonant error rate (CoER) and vowel error rate (VoER) of consonants and vowels extracted from the joint tiers of Models 1, 3, and 4, and from the phone tier of Model 2. These error rates are computed by deleting all out-of-class symbols from both the reference and hypothesis transcripts, and then computing the string edit distance between reference and hypothesis (for example, CoER is computed by deleting all non-consonant symbols from both reference and hypothesis).
This method usually gave PER and TER, for Models 3 and 4, that are worse than their  corresponding results in Table~\ref{tab:multitier}. In order to facilitate comparison between the tables, therefore, the lowest PER or TER in each column of Table~\ref{tab:multitier_breakdown} is bold {\bf only if} it is lower than the corresponding best entry in Table~\ref{tab:multitier}.  As shown in Table~\ref{tab:multitier_breakdown}, M1+F0 provides the lowest CoER and VoER in every language, but not always the best PER.  Closer study shows that, without F0 inputs, Model 1 always provides the best consonant error rates, but not always the best vowel error rates.  Tone behaves in a surprising manner.  Without F0, none of the TER entries in Table~\ref{tab:multitier_breakdown} are lower than Table~\ref{tab:multitier}.  Even with F0, the M1+F0 entry in Table~\ref{tab:multitier_breakdown} beats that of Table~\ref{tab:multitier} for only one language.  We conclude tentatively that consonants and vowels are best recognized using an output tier that requires them to carry their tone markings (Model 1), but that tone is best recognized using a separate output tone tier (Model 4 in the Multilingual case, Model 2 in the case of Lao).

\section{Conclusions}
\label{conclusions}
This experiment compared four methods for Multilingual and Cross-lingual CTC ASR of tones and phones.
Cross-lingual results must be considered tentative, because only one language (Lao) was available as the target of Cross-lingual ASR; future work should repeat the Cross-lingual experiment for all four languages (or more), using a cross-validation training paradigm. Nevertheless, some results of this experiment seem very clear, and likely to be supported by future experimentation.
Both synchronous (Model 1) and asynchronous (Models 2, 3, and 4) phones and tones can be adapted Cross-lingually, resulting in error rates far below those achieved by a Monolingual system trained on the same limited data.  An output tier that requires tone-marking of every vowel results in lower joint error rates, as well as lower error rates for both consonants and vowels separately, than the systems that recognize phones and tones on separate output tiers.  Conversely, tones are most accurately recognized using a system with separate phone and tone output tiers. The lowest tone error rates in the Multilingual case are provided by a multitask system with four output tiers (phone, tone, voice quality, and joint), while the lowest tone error rate for Cross-lingual ASR is provided by a system with two output tiers (phones and tones).

\bibliographystyle{IEEEtran}

\bibliography{mybib}

\end{document}